\documentclass[a4paper, 10pt]{article}

\usepackage{osameet2}
\usepackage{setspace} 

\usepackage[acronym,nomain]{glossaries}
\usepackage{amsmath,amssymb}
\usepackage{textgreek}
\usepackage{xfrac}
\usepackage{subfig}    
\usepackage{enumitem,wrapfig}
\usepackage{pgfplots,tikz}
\usetikzlibrary{shapes}

\definecolor{byzantine}{rgb}{0.74, 0.2, 0.64}

\usepackage[pdftex,colorlinks=true,bookmarks=false,citecolor=blue,urlcolor=blue]{hyperref} 

\begin{document}



\title{Improving HD-FEC decoding via bit marking}
\author{
Alex~Alvarado\textsuperscript{(1)}, Gabriele Liga\textsuperscript{(1)}, Yi Lei\textsuperscript{(2)}, Bin Chen\textsuperscript{(1)}, and~Alexios Balatsoukas-Stimming\textsuperscript{(3)}}
\address{
  \textsuperscript{(1)}Department of Electrical Engineering, Eindhoven University of Technology, 
  The Netherlands.\\  
  \textsuperscript{(2)}School of Computer Science and Information Engineering, Hefei University of Technology, China.\\ 
  \textsuperscript{(3)}Telecommunications Circuits Laboratory, \'Ecole Polytechnique F\'ed\'erale de Lausanne, Switzerland. \\ 
 \vspace{-2mm}
  \email{a.alvarado@tue.nl}}

\begin{abstract}
We review the recently introduced soft-aided bit-marking (SABM) algorithm and its suitability for product codes. Some aspects of the implementation of the SABM algorithm are discussed. The influence of suboptimal channel soft information is also analyzed.
\end{abstract}

\vspace{-2mm}

\section{Introduction}

Forward error correction (FEC) is required in optical communication systems to meet the ever-increasing data demands. FEC codes that can boost the net coding gain (NCG) are of key importance, specially in systems targeting data rates of $400$~Gb/s and beyond. Soft-decision (SD) FEC codes provide large NCGs, however, they are not the best candidates for very high data rate applications due to their high power consumption and decoding latency. In  this context, simple but powerful hard-decision (HD) FEC codes are a promising alternative. 

HD-FEC codes providing large NCGs include staircase codes (SCCs) and product codes (PCs). PCs (and generalisations thereof) offer, at high code rates, a very good compromise between complexity and performance. This is due to the fact that PCs enable fast decoding of long codewords through the use of iterative algebraic decoding. Iterative bounded distance decoding (iBDD) is typically used, where rows and colums are iteratively decoded.


An emergent area of research for next generation high-speed fiber optics is the used of low-complexity HD-FEC \emph{aided} by soft-information. The general idea is to recover some of the loss caused by the use of a (low-complexity) HD decoder by using some soft information available from the channel. Examples of this approach include \cite{Hager_ECOC2017,Sheikh_ECOC2018,Yi_ISTC2018,Yi_arxiv}.  In this paper, we review one of these algorithms: the recently proposed soft-aided bit-marking (SABM) algorithm \cite{Yi_ISTC2018,Yi_arxiv}. We put special attention to its performance for PCs as well as the impact of suboptimal calculation of the soft information from the channel.

\section{The SABM Algorithm}

The SABM algorithm was first proposed in \cite{Yi_ISTC2018} and later studied in detail in \cite{Yi_arxiv}. In what follows, we first review the SABM algorithm and then discuss some aspects of its implementation. We start by introducing an important quantity for this algorithm, the LLRs, which represent the soft information for bit $k$ of a channel observation $y_l$. For a complex AWGN channel withs SNR $\rho$, LLRs are calculated as
\begin{equation}\label{LLR}
   \lambda_{l,k}=\sum_{b \in \{0,1\}} (-1)^{1-b} \log\sum_{x_i \in \mathcal{X}_{k,b}} \textrm{exp}\left(-\frac{|y_{l}-\sqrt{\rho}x_{i}|^{2}}{2}\right)\approx\sum_{b \in \{0,1\}} (-1)^{b} \min_{x_i \in \mathcal{X}_{k,b}} |y_{l}-\sqrt{\rho}x_{i}|^2,
\end{equation}
where the set $\mathcal{X}_{k,b}$ enumerates all the constellation points in $\mathcal{X}$ whose $k$th bit is $b$. In hardware implementations, the much simpler max-log approximation is used, which is shown in the r.h.s. of \eqref{LLR}.

\subsection{Review of the Algorithm}

\begin{figure}[tb]
 \centering
 \raisebox{5ex}{\includegraphics[width=0.45\textwidth]{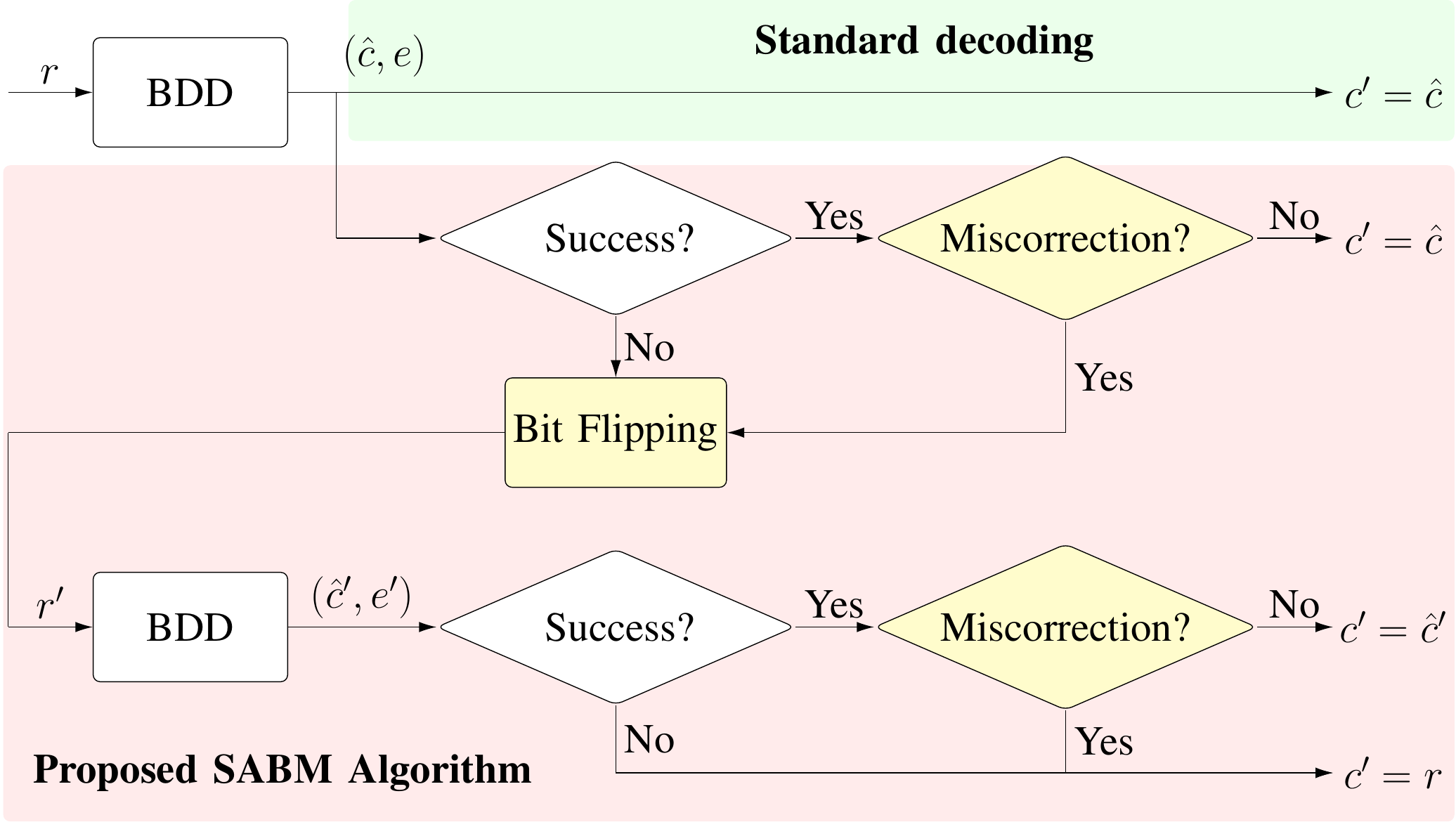}}
\hspace{3ex}
\resizebox{0.40\textwidth}{!}{\begin{tikzpicture}
\begin{axis}[%
every axis/.append style={font=\normalsize},
width=110mm,
height=0.47\textwidth,
xmin=6.7,
xmax=7.6,
xtick={6.3,6.5,...,7.5},
xmajorgrids,
xlabel={SNR (dB)},
ymin=2e-10,
ymax=3e-2,
ytick={1e-9,1e-8,1e-7,1e-6,1e-5,1e-4,1e-3,1e-2,1e-1},
ymajorgrids,
ymode=log,
ylabel= {Post-SCC BER},
legend style = {text=black,legend columns=3,legend cell align=left, at={(0.5,1.17)},anchor=north,font=\footnotesize}
]

\addplot [color=blue!95!black,mark=square*,mark size=1.8pt, thick,mark options={fill=white,solid},thick]
  table[row sep=crcr]{%
6.75000000000000	0.00990347896516655\\
6.80000000000000	0.00823318133862548 \\
6.85000000000000	0.00638570139633243\\
6.90000000000000	0.00354844395229213 \\
6.95000000000000	0.000649668875991850\\
7	3.40159243740600e-05\\
7.05000000000000	1.20503795364346e-06\\
7.10000000000000	4.56538209783399e-08\\
7.15000000000000	5.02409018443678e-09\\
7.20000000000000	2.03564026784816e-09\\
7.25000000000000	8.87991724076626e-10\\
  };
\addlegendentry{SABM Algorithm};

\addplot [color=green!95!black,mark=star,mark size=1.8pt, thick,mark options={fill=white,solid},thick]
  table[row sep=crcr]{%
6.80000000000000	0.00865784063823228 \\
6.85000000000000	0.00741863822633457 \\
6.90000000000000	0.00534824889087804\\
6.95000000000000	0.00171305977033612\\
7	                0.000123211218208068\\
7.05000000000000	6.07843501380837e-06\\
7.10000000000000	1.46245783368278e-07\\
7.15000000000000	9.45549315487450e-09\\
7.20000000000000	1.31313594385880e-09\\
7.25000000000000    6.97796154515846e-10\\
  };
\addlegendentry{Miscorrection-free decoding};

\addplot [color=magenta!95!black,mark=otimes*,mark size=1.8pt, thick,mark options={fill=white,solid},thick]
  table[row sep=crcr]{%
6.91879049527974  0.00834900000000000	 \\
6.96411467414104  0.00609700000000000\\
7.01039792197892  0.00276100000000000\\
7.05769542522044  0.000387300000000000\\
7.10606745793397  2.04000000000000e-05\\
7.15558004036256  5.63000000000000e-07	\\
7.20630570889022 1.13000000000000e-08	\\
7.25832442104719 9.87000000000000e-10	\\
7.31172462524322 6.05000000000000e-10	\\
  };
\addlegendentry{Method in [1]};

\addplot [color=brown!95!black,mark=diamond*,mark size=2.0pt, thick,mark options={fill=white,solid},thick]
  table[row sep=crcr]{%
7	0.00834951628670129
7.05000000000000	0.00692947571228756 \\
7.10000000000000	0.00507539117235673 \\
7.15000000000000	0.00245344755267057\\
7.20000000000000	0.000430676915740964\\
7.25000000000000	3.48941644819885e-05\\
7.30000000000000	1.53000000000000e-06\\
7.35000000000000	6.05865824880391e-08\\
7.40000000000000	8.39167908052808e-09\\
7.45000000000000	2.17048153105142e-09\\
7.47500000000000    1.84836847075853e-9\\
  };
\addlegendentry{Method in [6]};

\addplot [color=red!95!black,mark=*,mark size=1.8pt, thick,mark options={fill=white,solid},thick]
  table[row sep=crcr]{%
7.05000000000000	0.00744345895739743 \\
7.10000000000000	0.00601481917433723 \\
7.15000000000000	0.00401866681134846 \\
7.20000000000000	0.00159523414261445 \\
7.25000000000000	0.000250320162515669 \\
7.30000000000000	1.62988947351598e-05 \\
7.35000000000000	8.53297939527222e-07 \\
7.40000000000000	4.51708906359513e-08 \\
7.45000000000000	3.42986888565599e-09 \\
7.50000000000000    2.919567670211782e-09\\
  };
\addlegendentry{Standard decoding};



\draw [black!95!black,<->, line width=0.2mm] (axis cs: 7.088,1e-7) -- node [midway,above] {}  (axis cs: 7.385,1e-7);
\node at(axis cs: 7,1e-7) {\small $0.30$~dB};

\end{axis}
\end{tikzpicture}
}
\vspace{-2ex}
\caption{\small Left: Schematic diagram of the SABM algorithm, where miscorrection detection and bit flipping is used (from \cite{Yi_arxiv}), and Right: Post-SCC BER vs. SNR for code rate $R=0.87$ and $2$-PAM (from \cite{Yi_arxiv}).} 
\label{fig1}
\end{figure}
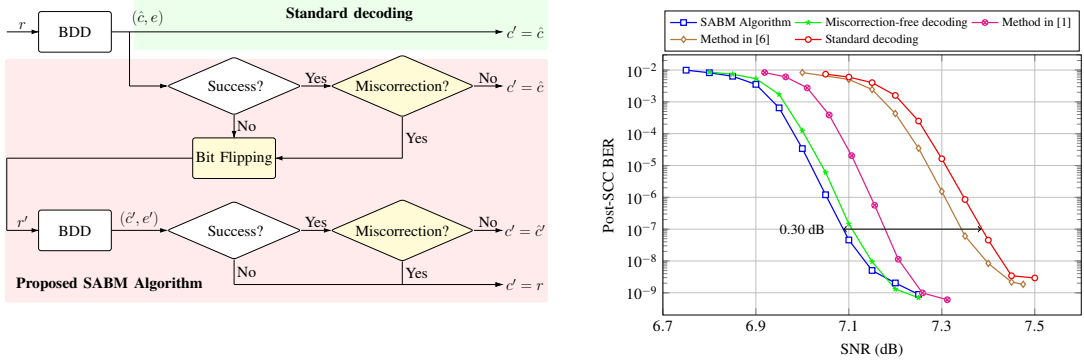

Fig.~\ref{fig1}~(left) shows a schematic diagram of the SABM algorithm (red area) as well as the standard BDD decoding (green area) used to decode a received sequence $r$. The SABM-based decoder marks some bits as highly reliable bits (HRBs) based on their LLRs, i.e., bits whose LLRs satisfy $|\lambda_{l,k}|>\delta$. The SABM algorithm also marks  $d_{0}-t-1$ highly unreliable bits (HUBs) with the smallest $|\lambda_{l,k}|$, where $d_0$ and $t$ are the minimum Hamming distance and error-correcting capability of the component code, respectively. Based on the marked bits, miscorrection detection (MD) is performed if BDD declares decoding success. If the flipped bits are HRBs \emph{or} are involved in zero-syndrome codewords, the output of BDD will be regarded as a miscorrection. Further, bit flipping (BF) is performed to handle those miscorrections and decoding failures. In the case of decoding failure, the most unreliable bit is flipped. In the case of misscorrection, the most $d_{0}-w_{\text{H}}(e)-1$ unreliable bits are flipped, where $e$ is the detected error pattern. The intuition here is that bits with the lowest reliability are the most likely channel errors. In some cases, BF will make the obtained sequence $r'$ close enough to the transmitted codeword. Thus, when the decoder calls BDD again, the remaining errors in $r'$ can be corrected. In case that BF results in a wrong decision, MD will make a final check if the decoding is successful.

Fig.~\ref{fig1} (right) shows standard results obtained for the SABM algorithm with $2$-PAM and a SCC with code rate of $R=0.87$. The component code we used is an extended Bose-Chaudhuri-Hocquenghem (BCH) code with parameters of $(256,239,2)$. As shown in Fig.~\ref{fig1}~(right), the BF operation makes the BER obtained with the SABM algorithm to be lower than the miscorrection-free case. The gain is $0.30$~dB at a bit error ratio (BER) of $10^{-7}$ when compared to standard decoding. Furthermore, it outperforms the previously proposed methods in \cite{Hager_ECOC2017, SmithPhD}. One important practical aspect to consider is that within each decoding window, only the soft information in the last block is used. Therefore, the SABM algorithm is only applied to the decoding related to the last block. 

\subsection{Practical Implementation Aspects}

The additional complexity incurred by the BFs in the SABM algorithm was analyzed in terms of the average number of calls to the component BDD decoder in \cite[Sec.~IV-C]{Yi_arxiv}. The relative complexity increase caused by the SABM algorithm with respect to standard  decoding for SCCs (window size $L$ and $\ell$ iterations) is given by $\eta\triangleq\frac{\overline{N}-N_{\text{sd}}}{N_{\text{sd}}}=\frac{\overline{N}-w(L-1)\ell}{w(L-1)\ell}$, where $\overline{N}$ and $N_{\text{sd}}$ are the number of BDD calls for the SABM algorithm and for the standard decoding, resp. It was shown in \cite[Sec.~IV-C]{Yi_arxiv} that the relative complexity increase is negligible, especially for very low post-SCC BER values. However, there are several other complexity and implementation aspects that have to be examined, especially if the decoder is to be implemented in hardware and integrated into a larger system. 

One potential issue is that the per-codeword running time of the algorithm is not deterministic, i.e., it is more difficult to guarantee a minimum throughput. Since the optical link delivers codewords at a constant rate, a large number of incoming codewords may have to be stored temporarily in a buffer while the SABM algorithm is still busy decoding the previous codeword. If several codewords that require multiple bit-flips to be decoded arrive consecutively, the buffer may overflow, resulting in dropped codewords which effectively increase the BER of the decoder. A potential solution is to employ a scheduler that adapts the maximum number of allowed bit-flips depending on the available space in the buffer.

The SABM algorithm also requires identifying a relatively small number of HUBs with the smallest absolute LLR value. There are two main problems with this task. First, even though only a small number of HUBs needs to be identified, the total number of HUBs is potentially very large, which increases the complexity of the sorting process.. In the worst case where the entire codeword is unreliable, there can be $2w^2$ HUBs. Second, the total number of HUBs is not constant, further complicating a potential hardware implementation which typically has to be provisioned for the worst case.

Another interesting aspect of the SABM algorithm is that the \emph{flipping mask} (i.e., the locations of HRBs and HUBs) within the last block of size $w \times w$ needs to be stored. If this mask is sparse, sparse matrix storage techniques can be exploited in order to minimize the storage space that is required for the flipping mask.


\section{Numerical Results}
In this section, we show numerical results on the performance of PC with a standard iterative BDD and with SABM decoding. The  PC uses as a component code an 1 bit-extended BCH code with parameters $(128,113,2)$ and rate $R=0.88$ (overall rate $R=0.78$ for the PC). Both decoders use 10 decoding iterations. The parameters of the SABM, such as reliability threshold $\delta$ and the number of miscorrection detection iterations are optimised to achieve the best post-FEC BER for this specific BCH code. It is found that $\delta=5$ and performing miscorrection detection for 5 iterations (out of 10) results in an optimal performance. The LLRs are calculated using both an exact expression and max-log approximation in \eqref{LLR}. When $4$-PAM and $8$-PAM modulation formats are evaluated, a random interleaver is used to ensure errors are independent are equally distributed across each PC block.  

\begin{figure}[!t]
\centering
\input{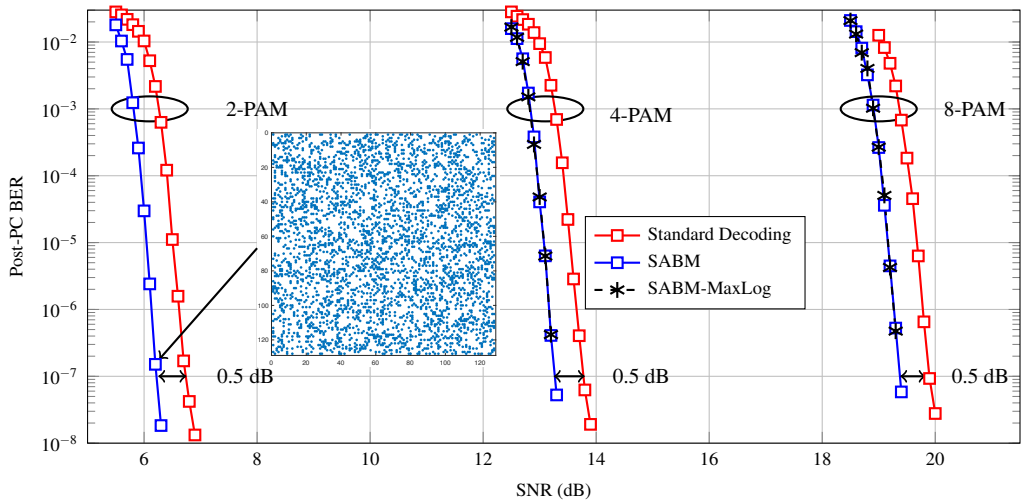}
\vspace{-2ex}
\caption{\small Post-PC BER vs. SNR for $2$-PAM, $4$-PAM and $8$-PAM and code rate $R=0.78$. The inset shows the flipping mask.}
\label{fig:PC}
\end{figure}

Fig.~\ref{fig:PC} shows the BER after decoding as a function of SNR in an AWGN channel, using both decoders, and for different modulation formats ($2$-PAM, $4$-PAM and $8$-PAM). The results show a significant coding gain for SABM over the standard decoding ($0.5$~dB at a BER of $10^{-7}$). This gain is preserved across all the different modulation formats shown. It can be seen that the standard decoder starts to floor in the region between $10^{-7}$ and $10^{-8}$. On the contrary, error floors are not observed for the SABM algorithm down to a BER of $10^{-8}$. 

The inset in Fig.~\ref{fig:PC} shows the positions where bits were \emph{not} marked as HRBs for an SNR of $6.2$~dB. The total number of marked positions was 3361 (out of $128\times 128$). For higher BERs, this number increases. For example, for $5.8$~dB this number was 4077. For $2$-PAM, the marking \emph{ratio} is therefore approximately in between $20$ and $25$~\%. Slightly smaller values were observed for $4$-PAM and $8$-PAM.

All of the above discussed results are obtained using for the SABM decoder a demapper which exactly calculates the LLRs. However, it can be also seen that using a max-log approximation for the $4$-PAM and $8$-PAM cases (stars) does not lead to any performance degradation of the SABM decoder. This is explained by the fact that the LLR approximation is particularly inaccurate for values of the LLRs above the threshold used for SABM, therefore not impacting the list of HUBs (bits with $|\lambda_{l,k}|\leq$5) and their sorted order.


\section{Conclusions}

In this paper we reviewed a new soft-decision aided decoding algorithm for hard-decision forward error correcting codes. Some practical implementation aspects of the algorithm were discussed. We believe this algorithm has potential to be used in 400G and beyond fiber optical systems.

\vspace{1ex}
\footnotesize
{\setstretch{0.02} 
\noindent \textbf{Acknowledgements:} The work of A. Alvarado and G. Liga is supported by the Netherlands Organisation for Scientific Research (NWO) via the VIDI Grant ICONIC (project number 15685) and has received funding from the European Research Council (ERC) under the European Union’s Horizon 2020 research and innovation programme (grant agreement No 757791). The work of G. Liga is also supported by NWO Visitor’s Travel Grant 040.11.659/6291.}

\vspace{-0mm}
\bibliographystyle{osajnl}

\let\oldthebibliography=\thebibliography
  \let\endoldthebibliography=\endthebibliography
  \renewenvironment{thebibliography}[1]%
  { \renewcommand{\baselinestretch}{0.98}%
    \begin{oldthebibliography}{#1}%
    \setlength{\parskip}{0pt}
    \setlength{\itemsep}{0pt}%
  }%
  {\end{oldthebibliography}}

\end{document}